# Results and perspectives of the solar neutrino experiment Borexino

G. Ranucci,[*,1] G. Bellini,[1] J. Benziger,[2] S. Bonetti,[1] M. Buizza Avanzini,[1] B. Caccianiga,[1] L. Cadonati,[3] F. Calaprice,[4] C. Carraro,[5] A. Chavarria,[4] F. Dalnoki-Veress,[4] D. D'Angelo,[1] H. de Kerret,[6] A. Derbin,[7] A. Etenko,[8] K. Fomenko,[9] D. Franco,[1] C. Galbiati,[4] S. Gazzana,[10] M. Giammarchi,[1] M. Goeger-Neff,[11] A. Goretti,[4] C. Grieb,[12] S. Hardy,[12] Aldo Ianni,[10] Andrea Ianni,[4] M. Joyce,[12] V. Kobychev,[13] G. Korga,[10] D. Kryn,[6] M. Laubenstein,[10] M. Leung,[4] T. Lewke,[11] E. Litvinovich,[8] B. Loer,[4] P. Lombardi,[1] L. Ludhova,[1] I. Machulin,[8] S. Manecki,[12] W. Maneschg,[14] G. Manuzio,[5] F. Masetti,[15] K. McCarty,[4] Q. Meindl,[11] E. Meroni,[1] L. Miramonti,[1] M. Misiaszek,[16, 10] D. Montanari,[10, 4] V. Muratova,[7] L.Oberauer,[11] M. Obolensky,[6] F. Ortica,[15] M. Pallavicini,[5] L. Papp,[10] L. Perasso,[1] S. Perasso,[5] A. Pocar,[4,] R.S. Raghavan,[12] A. Razeto,[10] P. Risso,[5] A. Romani,[15] D. Rountree,[12] A. Sabelnikov,[8] R. Saldanha,[4] C. Salvo,[5] S. Schönert,[14] H. Simgen,[14] M. Skorokhvatov,[8] O. Smirnov,[9] A. Sotnikov,[9] S. Sukhotin,[8] Y. Suvorov,[1, 8] R. Tartaglia,[10] G. Testera,[5] D. Vignaud,[6] R.B. Vogelaar,[12] F. von Feilitzsch,[11] M. Wojcik,[16] M. Wurm,[11] O. Zaimidoroga,[9] S. Zavatarelli,[5] and G. Zuzel [14]

(Borexino Collaboration)

[*]Contribution presented by G. Ranucci at ICHEP08. E-mail: ranucci@mi.infn.it

[1]Dipartimento di Fisica, Università degli Studi e INFN, 20133 Milano, Italy

[2]Chemical Engineering Department, Princeton University, Princeton, NJ 08544, USA

[3]Physics Department, University of Massachusetts, Amherst, MA 01003, USA

[4]Physics Department, Princeton University, Princeton, NJ 08544, USA

[5]Dipartimento di Fisica, Universit`a e INFN, Genova 16146, Italy

[6]Laboratoire AstroParticule et Cosmologie, 75231 Paris cedex 13, France

[7]St. Petersburg Nuclear Physics Institute, 188350 Gatchina, Russia

[8]RRC Kurchatov Institute, 123182 Moscow, Russia

[9]Joint Institute for Nuclear Research, 141980 Dubna, Russia

[10]INFN Laboratori Nazionali del Gran Sasso, SS 17 bis Km 18+910, 67010 Assergi (AQ), Italy

[11]Physik Department, Technische Universität Muenchen, 85747 Garching, Germany

[12]Physics Department, Virginia Polytechnic Institute and State University, Blacksburg, VA 24061, USA

[13]Kiev Institute for Nuclear Research, 06380 Kiev, Ukraine

[14]Max-Planck-Institut für Kernphysik, 69029 Heidelberg, Germany

[15]Dipartimento di Chimica, Universit`a e INFN, 06123 Perugia, Italy

[16]M. Smoluchowski Institute of Physics, Jagiellonian University, 30059 Krakow, Poland

Borexino is a massive, calorimetric, liquid scintillator detector aimed at the detection of low energy sub-MeV solar neutrinos, installed at the Gran Sasso Laboratory. After several years of construction, data taking started in May 2007, providing immediately incontrovertible evidence of the unprecedented radiopurity of the target mass, at the level required to ensure the successful detection of $^7$Be solar neutrinos, which was then announced in the 2007 summer. In this talk first the main technical characteristics of the detector will be highlighted, with special emphasis on the exceptional purity challenges successfully faced by the Collaboration, and afterwards the physics outputs reached so far will be carefully reported and illustrated, together with the perspectives for the future measurements that will complete the broad program of the experiment.

## 1. INTRODUCTION

The Borexino Collaboration reported in early works [1], [2] the first real time measurement of the flux of the monoenergetic neutrinos from the $^7$Be electron capture in the core of the Sun, marking a fundamental scientific and technological breakthrough in the experimental field of solar neutrinos. This success represents the achievement of a 20 year long research, deeply rooted in the quest and development of purification techniques able to reach the new ground of the unprecedented radiopurity levels required for the successful $^7$Be neutrino detection. In particular, the impressive background results obtained through the pilot prototype CTF [3], the first demonstration ever obtained of scintillator purities in the range required for low energy solar neutrino spectroscopy, were the key which opened the way towards the Borexino challenge.

The present work reports an account of the $^7$Be neutrinos flux measurement, based upon almost 200 days of data, after a short description of the main features of the detector and of the unprecedented low background levels achieved. Finally the future, further measurement perspectives of the experiment are described.

## 2. DETECTOR DESCRIPTION

The key features of the Borexino detector and of its components have been thoroughly described in [4][5], and thus only a succinct summary is reported here.

Borexino is a scintillator detector which employs as active detection medium a mixture of pseudocumene (PC, 1,2,4-trimethylbenzene) and PPO (2,5-diphenyloxazole, a fluorescent dye) at a concentration of 1.5 g/l. Because of its intrinsic high luminosity (50 times more than in the ˇCerenkov technique) the liquid scintillation technology is extremely suitable for massive calorimetric low energy spectroscopy. However, no directionality is possible and therefore, as a consequence, it is not possible to distinguish neutrino scattered electrons from electrons due to natural radioactivity. Thus the key requirement in the technology of Borexino is an extremely low radioactive contamination.

To reach ultra low operating background conditions in the detector, the design of Borexino is based on the principle of graded shielding, with the inner core scintillator at the center of a set of concentric shells of increasing radiopurity. The scintillator mass (278-ton) is contained in a 125 μm thick nylon Inner Vessel (IV) with a radius of 4.25 m. Within the IV a fiducial mass is software defined through the estimated events position, obtained from the PMTs timing data via a time-of-flight algorithm.

A second nylon outer vessel (OV) with radius 5.50 m surrounds the IV, acting as a barrier against radon and other background contaminations originating from outside. The region between the inner and outer vessels contains a passive shield composed of pseudocumene and 5.0 g/l DMP (dimethylphthalate), a material that quenches the residual scintillation of PC so that spectroscopic signals arise dominantly from the interior of the IV.

A 6.85 m radius stainless steel sphere (SSS) encloses the central part of the detector and serves also as a support structure for the 2212 8" PMTs (ETL 9351), each equipped with an aluminium light concentrator. The region between the OV and the SSS is filled with the same inert buffer fluid (PC plus DMP) which is contained between the inner and outer vessels. Finally, the entire detector is contained in a tank (radius 9 m, height 16.9 m) of ultra-pure water, in which the 200 PMT's forming the muon veto detector are immersed.

It must be also mentioned that for the success of the experiment key elements are the many liquid purification and handling systems, which were designed and installed to ensure the proper manipulation of the fluids at the exceptional purity level demanded by Borexino.

## 3. GENERAL DETECTOR FEATURES

### 3.1. Expected signals and backgrounds

In Borexino the extraction of the neutrino flux relies upon the precise fit of the data to a signal plus background model obtained through the weighted sum of the theoretical spectra of all the expected contributions.

The expected signals are clearly the neutrino spectra which are computed according to the standard solar model and the MSW-LMA [6] oscillation framework. Since the analysis window is up to 2 MeV, the spectra of interest are those of the pp, pep and $^7$Be neutrinos from the pp chain, as well as the spectrum stemming from the CNO cycle.

The other spectra of interest are those of the dominant backgrounds. Experimentally we observed that the prominent contaminants are, at low energy, the $^{14}$C, at high energy the $^{11}$C of cosmogenic origin, and at the energy of the $^7$Be recoil spectrum the $^{85}$Kr and the $^{210}$Po. Their respective spectra have been thus carefully computed for the purpose of neutrino analysis.

### 3.2. Events reconstruction and detector performances

Borexino is a self-triggering multiplicity detector, and thus the main trigger fires when at least K PMTs detect one photoelectron within a time window of 60 ns. Typically, K was equal to 30 in the data runs considered in this work, corresponding approximately to an energy threshold of 60 keV.

Upon triggering, time and charge of each firing PMT are acquired and stored. The time is measured by a Time to Digital Converter (TDC) with a resolution of about 0.5 ns, while the charge (after integration and shaping of the PMT anode pulses) is measured through an 8 bit Analog to Digital Converter (ADC).

The readout sequence can also be activated by the outer muon detector through a suitable triggering system, which fires when at least six PMTs detect light in a time window of 150 ns. Regardless of the trigger type, both data from the inner and outer detectors are always acquired. The typical triggering rate during the runs analyzed in this paper was 15 Hz, including all trigger types. This rate is largely dominated by very low energy $^{14}$C events.

Event positions are estimated by analysis of the times of the triggered PMTs via a time-of-flight based likelihood methodology. The energy resolution scales as $5\%/\sqrt{E(MeV)}$, while the position resolution is about 40 cm @ 150 keV.

The estimated light output is about 500 pe/MeV.

### 3.3. Measured background levels

As somehow expected from the successful CTF experience, U and Th proved immediately after the data taking start-up to be at an extremely low concentration level, i.e. $(1.6\pm0.1) \cdot 10^{-17}$ g/g for U and $(6.8\pm1.3) \cdot 10^{-18}$ g/g for Th. For natural K the remarkable upper limit of $<3 \cdot 10^{-14}$ g/g has been obtained, as well.

Other important backgrounds in sizable, though tolerable, amount are $^{85}$Kr, evaluated to be $29\pm14$ counts/(day·100 ton) via a special delayed coincidence tag, and $^{210}$Po (initial contamination of about 80 counts/day/ton, decaying afterwards following the intrinsic 200 days lifetime). By far, however, the major contaminant is $^{14}$C, whose isotopic ratio is evaluated from the detected counting rate equal to $^{14}$C/$^{12}$C$=(2.7\pm0.6) \cdot 10^{-18}$, perfectly suited for the planned analysis threshold of 200 keV. To be noted, finally, also the presence of the cosmogenic $^{11}$C signals, observed at an average rate of 25 c/d/100t, which is the range of the predictions stemming from the previous studies reported in [7] and [8], though slightly higher.

With the exception of the first couple of months after the end of the fill, the detected Rn contamination is very limited, at the level of few counts per week, the emanation from the vessel being its major source throughout most of the data taking period.

### 3.4. Event selections

The results presented here concerns 192 live days between May 2007 and February 2008. Event selection is performed according to the following criteria:

(i) Only single cluster events are accepted, so to exclude pile-up and fast coincident events.

(ii) Muon events are rejected by means of the muon flag, i.e. of the signal registered in the outer water tank detector.

(iii) After each muon crossing the scintillator, all events (afterpulses and spurious events) within a time window of 2 ms are rejected.

(iv) The Radon induced $^{214}$Bi-$^{214}$Po sequences are identified and removed, as well as their precursor $^{214}$Pb signals.

(v) In order to remove the overwhelming external background, for neutrino analysis only signals reconstructed within a spherical 100 t fiducial volume are accepted.

Fig. 1 displays the spectra resulting from the selection cuts starting from the total raw spectrum. The solid black curve is the initial spectrum with only cuts (i-iii) applied. Two prominent components, $^{14}$C below 80 pe (i.e. photoelectrons) and $^{210}$Po at about 190 pe, are immediately visible.

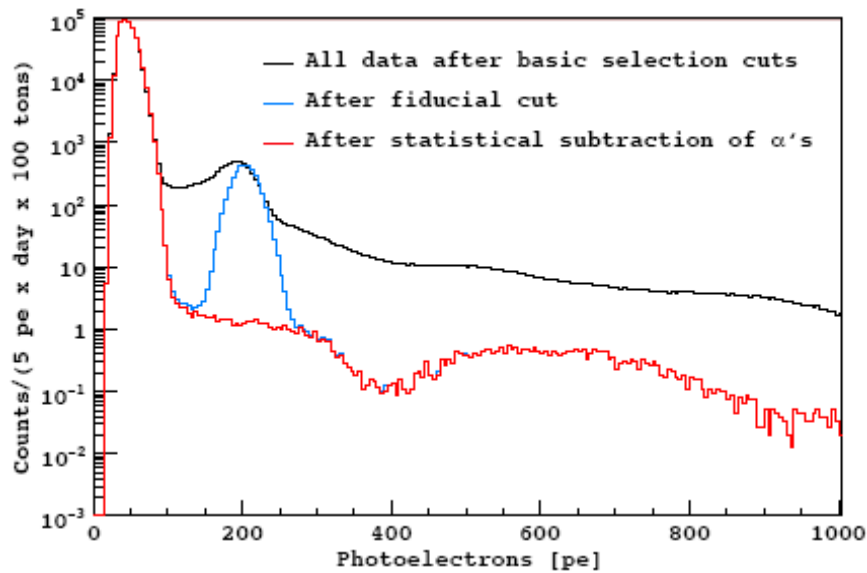

Figure 1 : The raw photoelectron charge spectrum after the basic cuts i–iii (black), after the Radon and fiducial volume cuts iv- v (blue), and after the statistical subtraction of the α-emitting contaminants (red).

The solid blue curve is the spectrum obtained after all of the above cuts (i-v) are applied, showing the dramatic effect linked to the removal of the external background. Now, in addition to $^{14}$C and $^{210}$Po, also the $^7$Be shoulder and the broad spectrum of $^{11}$C are clearly visible. Finally the red curve is the spectrum left over by the application of the alpha-beta discrimination technique, which allows a powerful rejection of the polonium peak.

## 4.  EXTRACTION OF THE $^7$BE SOLAR NEUTRINO FLUX

The final spectrum after all the cuts is fitted to a global signal-plus-background model to extract quantitatively the value of the $^7$Be flux. Two independent analyses, providing consistent results, have been carried out.

The fit is performed roughly from 160 keV, thus including the extreme tail of the $^{14}$C, to 2 MeV, encompassing the entire $^{11}$C region. Free parameters are, besides the light yield, the amplitude multiplicative factors of the $^7$Be, CNO, $^{85}$Kr, $^{14}$C and $^{11}$C spectra. The pp and pep spectra are included in the fit, but at their nominal model value.

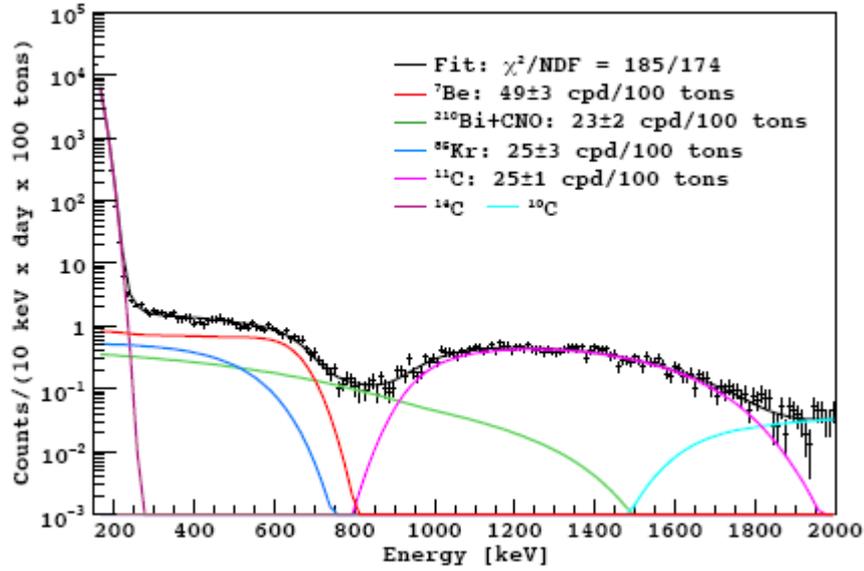

Figure 2. Global spectral fit in the energy region 160–2000 keV.

The fit output for the 192 days data sample considered here is reported in Fig. 2; the results are conventionally expressed in counts/day/100 tons of scintillator : the $^7$Be count rate is estimated to be 49 ± 3, Kr is 25 ± 3, consistent with the estimate stemming from the coincidence analysis, cumulatively CNO and $^{210}$Bi, that at this stage cannot be disentangled, are 23 ± 2, and $^{11}$C is 25 ± 1. The errors quoted so far are only the statistical errors, on top of which also the systematic errors should be considered. The uncertainties in the fiducial volume and energy scale determinations are the major sources of systematic errors, originating in the end the final $^7$Be evaluation of 49 ± 3$_{stat}$± 4$_{sys}$ counts/day/100 tons, which translates into a $^7$Be flux of (5.08 ± 0.25) ·10$^9$ cm$^{-2}$s$^{-1}$, very well in agreement with the prediction of the BS07(GS98) Standard Solar Model [9]. For comparison, the detected count rate in case of absence of oscillations would have been 74 ± 4 counts/day/100 tons. The resulting electrons survival probability at the $^7$Be energy is $P_{ee}$=0.56± 0.10.

Therefore, Borexino on one hand spectacularly confirms the MSW-LMA solar neutrino oscillation scenario, and on the other provides the first direct measurement of the survival probability in the low energy vacuum MSW regime [10].

## 5. FUTURE PERSPECTIVES

Given the exceptional, unprecedented purity results achieved in Borexino, further measurements beyond the original goal of the $^7$Be detection are prospectively possible in the next years of running of the detector.

First of all a broad and accurate investigation of the solar neutrino spectrum is well within our experimental reach: not only the $^7$Be can be pin pointed to an accuracy of 5% (with respect to the 10% uncertainty of the measurement reported here), but also the other important medium and high energy components of the solar neutrino spectrum are suitable to be searched for. This is specifically true for the $^8$B neutrinos [11], as well as for the extremely challenging pep and CNO fluxes. For the latter two, in particular, it will be needed to cope with the background represented by the $^{11}$C signals, adhering to the strategy already devised by the Collaboration in [8].

The extremely low $^{14}$C level, coupled to the good achieved energy resolution, opens also a possible exploration window between 200 and 240 keV in which the observation of the fundamental pp flux can be attempted.

Other important fields of investigation for Borexino will be the neutrino magnetic moment (a limit of < 5.4·10$^{-11}$ $\mu_B$ (90% C.L.) being reported in [2]) , the geoneutrinos [12] and the supernovae search.

## 6. CONCLUSIONS

After a long quest for the ultimate radiopurity, Borexino entered with great success the solar neutrino experimental arena providing the first real time detection of $^7$Be solar neutrinos, marking a fundamental milestone in the field of ultra low background techniques, as well as in the solar neutrino physics.

Such a success opens the way to further investigations of other solar ν sources ($^8$B, pep, CNO, pp), being the rich Borexino program suitable to be extended to neutrino magnetic moment, antineutrinos and supernovae studies.